\documentclass[usenatbib]{mn2e}
\usepackage{graphicx}
\usepackage{ifthen}

\def\ltsima{$\; \buildrel < \over \sim \;$}
\def\simlt{\lower.5ex\hbox{\ltsima}}
\def\gtsima{$\; \buildrel > \over \sim \;$}
\def\simgt{\lower.5ex\hbox{\gtsima}}
\def\kms{{\rm\,km\,s^{-1}}}

\def\kpc{{\rm\,kpc}}

\def\AA{$\; \buildrel \circ \over {\rm A}$}

\def\UseFigs{1}

\def\deg{^\circ}

\def\s{\ifmmode \widetilde \else \~\fi}
\def\={\overline}

\def\spose#1{\hbox to 0pt{#1\hss}}

\def\lta{\mathrel{\spose{\lower 3pt\hbox{$\mathchar"218$}}
     \raise 2.0pt\hbox{$\mathchar"13C$}}}
\def\gta{\mathrel{\spose{\lower 3pt\hbox{$\mathchar"218$}}
     \raise 2.0pt\hbox{$\mathchar"13E$}}}
\def\Dt{\spose{\raise 1.5ex\hbox{\hskip3pt$\mathchar"201$}}}    
\def\dt{\spose{\raise 1.0ex\hbox{\hskip2pt$\mathchar"201$}}}    

\def\dotsfill{\leaders\hbox to 1em{\hss.\hss}\hfill}

\def\Gyr{{\rm\,Gyr}}

\loadboldmathitalic 
\title[Why the Canis Major overdensity is not due to the Warp] {Why the Canis Major overdensity is not due to the Warp:
analysis of its radial profile and velocities}
\author[N. F. Martin et al.] {N. F. Martin$^{1}$, R. A. Ibata$^{1}$, B. C. Conn$^{2}$, G. F. Lewis$^{2}$ , M.
Bellazzini$^{3}$, 
\newauthor M. J. Irwin$^{4}$,  \& A. W. McConnachie$^{4}$\\
$^{1}$ Observatoire de Strasbourg, 11, rue de l'Universit\'e, F-67000, Strasbourg, France\\
$^{2}$ Institute of Astronomy, School of Physics, A29, University of Sydney, NSW 2006, Australia\\
$^{3}$ INAF - Osservatorio Astronomico di Bologna, Via Ranzani 1, 40127, Bologna, Italy\\
$^{4}$ Institute of Astronomy, Madingley Road, Cambridge, CB3 0HA, U.K.\\
}

\date{\today}
\begin{document} 
\maketitle 
\begin{abstract} 
In    response   to    criticism   by    \citet{momany},    that   the recently-identified  Canis  Major (CMa) 
overdensity  could be  simply explained by the Galactic warp, we present proof of the existence of a stellar population
in the direction of CMa that cannot be explained by known Galactic  components.  By  analyzing the radial  distribution
of counts  of M-giant stars  in this  direction, we show  that the
\cite{momany}  warp model  overestimates the  number of  stars  in the Northern hemisphere, hence  hiding the CMa
feature in  the South.  The use  of  a better  model  of  the warp  has  little  influence on  the morphology of the
overdensity and clearly displays an excess of stars grouped at a distance of $D=7.2\pm 0.3 \kpc$.  To lend further
support to the existence of a population  that does not belong to the Galactic disc, we present radial  velocities of
M-giant stars  in the centre  of the CMa structure that  were obtained  with the 2dF  spectrograph at  the AAT. The
extra population shows a radial  velocity of $v_{r}=109\pm4\kms$, which is significantly higher than the typical
velocity of the disc at the  distance  of CMa.  This  population  also  has a  low  dispersion ($13\pm4\kms$).  The 
Canis  Major overdensity  is  therefore  highly unlikely to be due to the  Galactic warp, adding weight to the
hypothesis that we are observing a disrupting dwarf galaxy or its remnants. This leads to questions
 on what part of CMa  was previously  identified  as  the Warp and  how  to possibly disentangle the two structures.
\end{abstract}


\section{Introduction} 
The  recently-discovered Canis  Major overdensity  (hereafter CMa) appears  to  be an ongoing accretion  event that will
 contribute to the build-up  of the Galactic thick disk  \citep[][hereafter Paper~I]{martin}.  Found close to the plane
of the Milky Way disc, we argued this putative dwarf galaxy may be the progenitor of the `Ring' of stars that
encompasses the Galaxy \citep{ibata03,  crane}, which is  seen most  clearly in  the Galactic anticentre  direction
\citep{newberg,yanny}.  This `Ring'  of stars may  have been  built progressively  as stars  were removed  from that
dwarf galaxy by the disruptive tidal forces of the Milky Way. 

In Paper~I, we extracted candidate M-giant stars from the 2 Micron All Sky  Survey (2MASS), following  selection
criteria  used to  study the tidal  stream of  the Sagittarius  dwarf galaxy  \citep{majewski}.  By comparing the
M-giant distribution above and below the Galactic plane, we brought  to light several large-scale Galactic  asymmetries
that we interpreted  as the  CMa  dwarf galaxy  and  the CMa tidal stream.   \citet[][hereafter 
Paper~II]{bellazzini04}  presented  deep  photometry  of a field at $4.2\deg$ from the center of CMa  and of Galactic
open clusters that  lie fortuitously in front of the CMa overdensity.  These data give a good constraint on the distance
to this population, ${\rm (m-M)=14.6\pm0.3}$, and also constrain the dominant stellar population to be  of intermediate 
age ($\sim 4$--$10\Gyr$)  and to  be metal-rich ($-0.7\le [M/H]\le 0.0$).

However, \citet[][hereafter  M04]{momany} questioned the  existence of the CMa overdensity. They claimed that  it could
be entirely explained by a simple shift of the Galactic plane  of 2 degrees to the South to model the  Galactic  warp 
that is  known  to  exist  in  this part  of  the sky. Using  the UCAC2 proper  motion catalogue, they also  showed that
the M-giants  composing the overdensity are rotating  around the Milky Way  in  a  prograde manner  and  at  a 
tangential velocity  that  is compatible with the disc.

Here, we use the radial starcount distribution and the radial velocity of M-giants in  the direction of CMa to show  the
CMa overdensity {\it cannot}  be  explained  by  the  Warp  and  that  its  morphology  and kinematics are clearly
different from  what would be expected from the Warp. We refer the reader to Paper~II for a comparison of
observations in the CMa region with the Besan\c{c}on model of the Galaxy \citep{robin}.

Throughout this  work, we assume that  the Solar radius  is $R_\odot = 8\kpc$,  that the  LSR circular  velocity is 
$220\kms$, and  that the peculiar motion of the  Sun is ($U_0=10.00\kms, V_0=5.25\kms, W_0=7.17
\kms$; \citealt{dehnen98}).

\section{Radial distribution of Canis Major M-giants} 
To account  for M04's criticism  on the use of  the \citet[][hereafter S98]{schlegel} values for dust  extinction (which
have been claimed to overestimate the extinction in regions  of high reddening), we now use the \citet[][hereafter
B00]{bonifacio}  asymptotic correction of these and   redefine  our  sample   of  M-giants   (sample  A   of  Paper~I)
accordingly. The  magnitudes we consider  in this letter have  all been de-reddened in this way.

\subsection{Distances} 
In  Paper~I, we used  the Red  Giant Branch  of the  Sagittarius dwarf galaxy as a reference to  calculate the distances
to the M-giant stars in  our  sample  A.  As  we  already  noted,  the uncertainty  on  the metallicity and  age of the 
different stellar populations leads  to a possible $\sim  30\%$ uncertainty on the distances,  with the values obtained
being a lower limit.

Since the present argument on the existence of an overdensity in Canis
Major is  mainly based on  distance distributions, it is  desirable to
have an  independent and more  robust method to validate  our distance
estimates.  Therefore, we first apply  the Tip of the Red Giant Branch
(TRGB) algorithm of \citet{mcconnachie},  which they used to determine
the  distances  to the  M31  group of  dwarf  galaxies.   As has  been
explained  there, the  I-magnitude  has the  advantage  of being  only
slightly  dependent  on  metallicity.  It  is  therefore  particularly
adapted for the study of  the CMa population for which the metallicity
has not  yet been precisely  determined. However, to account  for this
lack of  precise metallicity, we  double the uncertainties  adopted by
\citet{mcconnachie} for  the I absolute magnitude of  the tip, leading
to ${\rm I}=-4.04\pm 0.10$ \citep{bellazzini01}.

Since the  2MASS catalogue we are  using only provides J,  H and ${\rm
K_s}$  magnitudes, we  cross identified  our sample  A with  the DENIS
catalogue that contains 190 million  objects down to I=18.5. With the
TRGB algorithm being sensitive to contamination from foreground stars, we
study  a large  area to  obtain sufficient  statistics. In  the region
$230\deg<l<250\deg$ and $-20\deg<b<-5\deg$,  3628 stars among the 6480
2MASS M-giant stars have their counterpart in DENIS and can be used to
determine the TRGB of the CMa population.

Applying  the  TRGB  algorithm  leads  to $i=10.25\pm  0.03$  for  the magnitude of the tip, which  corresponds to a
heliocentric distance of the  Canis  Major  population   of  $D=7.2\pm  0.3\kpc$.  This  is statistically  equivalent 
to the  previous values of $7.1\pm1.3\kpc$ of Paper~I and of $8\pm1\kpc$  of Paper~II  but with  a smaller 
uncertainty.

With this distance modulus, we are now able to obtain a photometric parallax to the CMa stars in the
same way \citet{majewski} did for the Sgr stars. Using the 2MASS colour-magnitude diagram of the same region, we compute
a linear fit to the Red Giant Branch (RGB) of the CMa population. As in \citet{majewski}, we restrict the fit to those
stars having $0.9<{\rm J-K_s}<1.1$ and apply a $2.5\sigma$ iterative rejection algorithm to discard the contaminating
disc stars. This leads to the following fit:
\begin{equation}
{\rm K_s}=-8.9({\rm J-K_s})+18.0
\end{equation}
with uncertainties slightly lower than 0.1 on the values. This result is close to the one \citet{majewski} deduced
for the Sagittarius dwarf, which explains the compatibility between the TRGB estimated distance to CMa and the 
estimate of Paper~I based on the Sgr fiducial.

Using a  TRGB algorithm and a  CMD fit of  the slope of the  Red Giant
Branch of  the Canis Major population,  we have derived  a relation to
estimate  the  distance  to   CMa  stars.  Contrary  to  our  previous
estimates, this relation does not require  the use of the Sgr RGB as a
reference and  should reduce the uncertainties  discussed in Paper~I.
Throughout this  letter, we will now  use relation (1)  to estimate the
distance to stars in the CMa region.

\subsection{The Canis Major overdensity is not the Warp} 
Momany  et  al.   (2004)  discarded  the  possibility  of  an  unknown population in  Canis Major, arguing that the 
overdensity presented in Paper~I  could be accounted  for by  correcting S98  extinction values with the asymptotic
correction  introduced by B00 and using $b=-2\deg$ as  the  symmetric plane  of  the  Galaxy to  model  the  Warp in 
the $235\deg<l<245\deg$ region.

Since  the CMa  overdensity appears  strong and  peaked in  the radial
distribution of M-giants in Paper  I, we checked the assumption of M04
by studying  the radial  distribution of M-giants  in the  same region
that they  used: $|b'|<20\deg$ and $235\deg<l<245\deg$,  where $b'$ is
the Galactic latitude calculated  from the warped Galactic plane. This
is shown on Figure~1 with the Galactic plane taken as $b=-2\deg$. Even
when using this  simple warp model of M04,  the radial distribution of
M-giants in the direction of CMa shows a clear overdensity of stars at
the  position  we  previously   identified  in  Paper  I  (centred  on
$D=7.2\kpc$).  Moreover,  the star  counts of the  Northern hemisphere
display an  asymmetric behaviour with an overabundance  of stars, with
respect to  the South, for  $D\lta5\kpc$ and $D\gta11\kpc$.   The star
counts should  be symmetric in  the lower distance interval  since the
Warp  only   begins  at  $D\sim   6\kpc$  in  the  direction   of  CMa
\citep[][hereafter  Y04]{yusifov}.   Furthermore, underestimating  the
displacement  of  the  Warp  at  large  distances  should  produce  an
overestimate of the counts in the {\it Southern} hemisphere for $D\gta
11\kpc$.   The  low  distance  overestimate  is to  be  expected  from
contamination by local  dwarfs; this is an artefact  of the simple M04
warp  model at short  distances, where  the warp  should not  have any
influence.  In the same way, fainter dwarfs that are contaminating the
sample are wrongly taken as M-giants at large distances.

\begin{figure}
\ifthenelse{\UseFigs=1}{
\includegraphics[angle=270,width=\hsize]{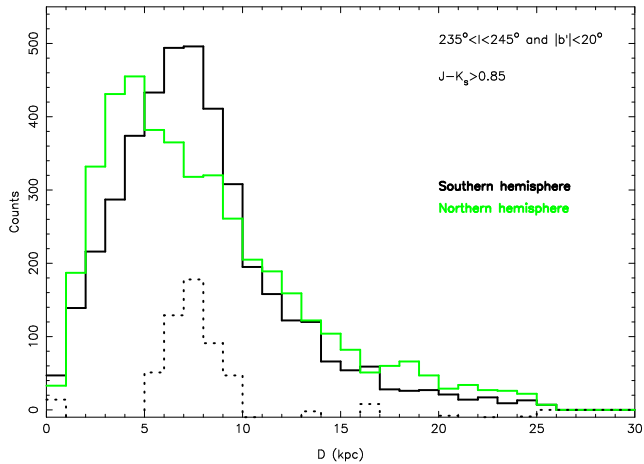} }{caption1}
\caption{Radial  distribution of  M-giant  stars above  and under  the warped Galactic plane for  $235\deg<l<245\deg$
and $|b'|<20$, with the Galactic  latitude  from the  warped  plane  $b'=b+2\deg$.  The  black histogram shows  the
distribution of Southern  ($b'<0\deg$) stars, the grey  histogram  that  of  Northern stars  ($b'>0\deg$).   The  dotted
histogram represents the  extra counts of stars in  the South compared to the North and is centered on 
$D\sim7$--$7.5\kpc$. The northern  hemisphere counts also  show a contamination by local stars that leads to an
overestimate of stars at short and long distances where the counts should be symmetric.}
\end{figure}

Due to the fact  that in M04, the star counts are  summed up along the
line of sight,  these overestimates of the Northern  counts {\it hide}
the  presence of  the  (Southern) CMa  overdensity  visible at  around
$7$--$7.5\kpc$.  However, since the angular maximum of the Southern warp is
thought  to  be  only  $30\deg$  away  (at  $l\sim270\deg$,  see  e.g.
\citealt{djorgovski};  \citealt{lopezcorredoira};  Y04), the  Galactic
warp should  indeed be  taken into account  when dealing with  the CMa
overdensity.  Therefore, we  re-analyze the  radial  distribution, but
this  time using  the  Y04 model  of the  warp.  To avoid  as much  as
possible contamination  of the M-giants  by local dwarfs, we  impose a
low latitude  cut at $|b'|=5\deg$ and  we restrict the  color range of
our  stars   to  $0.9<J-K_{s}<1.3$.  These   limits  are  particularly
important since extinction reaches high values near the plane and even
with  the B00  correction,  a small  underestimate  of the  extinction
shifts red disc  stars a little redder and they  can enter the M-giant
selection  box.  Shifting  the  lower  colour limit  to  the red  also
minimises this contamination.

\begin{figure}
\ifthenelse{\UseFigs=1}{
\includegraphics[angle=270,width=\hsize]{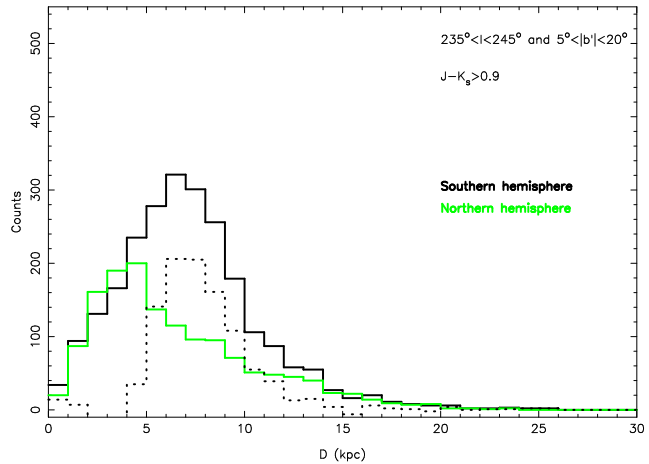} }{caption1}
\caption{Same as  Figure~1 but  using the more  precise warp  model of Y04 to obtain the  Galactic latitude from the
warped plane $b'$. To avoid contamination from  local stars, the sample is restricted to  $5\deg<|b'|<20\deg$ and  to
stars  having  $0.9<J-K_{s}<1.3$. Even with these  limits, the CMa overdensity is  clearer than  before with twice as
many stars as in Figure~1.}
\end{figure}

The results are  shown on Figure~2 and, as could  be expected from the
improvement over  the previous simple warp model,  the CMa overdensity
becomes more  clearly visible. Reducing our sample to
higher latitudes (e.g. $7\deg<|b'|<20\deg$) does not affect the distributions, meaning
that systematic selection effects produced be extinction are not responsible for the overdensity in the Southern
hemisphere. It can  also be noticed  that our more
conservative M-giant  selection and this  better modeling of  the Warp
corrects the asymmetry of  Figure~1 at $D\lta5\kpc$ and $D\gta11\kpc$,
which was caused  by overestimating the warping of  the Galactic plane
at low distances and from  asymmetric contamination of local dwarfs in
the high distance M-giant sample.   Even considering that the Y04 warp
model   does    not   have   as   large   a    displacement   as   the
\citet{lopezcorredoira} 2MASS-based model (see e.g.  Figure~2 of Y04),
it is  only at higher distances  than the bulk of  the CMa overdensity
that the two models actually deviate from each other. It could also be
argued that we are in fact observing the south Warp curling back up to
the Galactic  plane at  the edge  of the disc  as it  is seen  for the
gaseous warp \citep{burton}. In this  case, we would indeed expect the
stars to pile up along the line  of sight when the Warp returns to the
mean plane. However, this happens for the gas at much higher distances
($D_{GC}\sim18\kpc$)
and at  the estimated CMa distance,  the gaseous warp  gently dives in
the Southern hemisphere (as modeled  here). Therefore, a model that is
highly different  from what is currently known  of the stellar/gaseous
warp  should be  summoned to  explain the  CMa overdensity  in  a Warp
scenario.

Thus,  we have  to  conclude  that  the CMa overdensity presented in Paper I {\it really is an unknown feature that
appears in addition to the Galactic warp}.

Using  the  warp  model  changes only slightly  the counts and morphology  of CMa compared to
Paper~I. Indeed, the number of M-giants that belong to the overdensity only drops by $\sim10\%$, the distance to the
structure is still in  the same range  and in good  agreement with the  TRGB result presented above,  and its  FWHM
remains the  same. This means  that at this location, the Warp is only contributing a minor number of M-giant stars
compared to the CMa overdensity.

\section{Radial velocities of Canis Major M-giants}

To study this  population and other low latitude  structures at higher longitude in  greater depth we  undertook a
series of  observations on the  nights  of  April  7-12,  2004  with  the  2-degree  field  (2dF) spectrograph at the
Anglo-Australian Telescope (AAT).  We employed two different   spectrograph   settings,  with   the   1200V  grating  
on spectrograph  1 (covering  4600--5600\AA\ at  1\AA/pixel) and with the 1200R grating on spectrograph 2 (covering
8000--9000\AA, also at  1\AA/pixel).  While  a detailed  analysis of  this dataset will be presented 
in forthcoming papers, we focus  here on the radial velocities of  stars selected to  be M-giant candidates  obtained
with spectrograph 1 in a field centered on the CMa overdensity.  These were chosen from the colour-magnitude region of
sample A: ${\rm 0.85 < J-K_{s} < 1.30}$, ${\rm 0.561 (J-K_{s})  + 0.22 < J-H  < 0.561 (J-K_{s}) +  0.36}$, and with
estimated distances (using the above calibration) in the range $4\kpc< D < 20 \kpc$.

The  spectra  were  extracted  using the  `2dfdr'  reduction  software
\citep{taylor}, but  then calibrated  in wavelength and  corrected for
contamination  from sky  emission  using algorithms  developed by  our
group.  Using Fourier cross-correlation methods, the velocities of the
stars  were calculated  by comparison  to a  range of  radial velocity
standard stars  of type  F--M (both dwarfs  and giants).  The velocity
value corresponding to the  template that best-matched the spectrum of
the survey  star was selected.  The typical  resulting radial velocity
uncertainty   of   well-measured  M-giant   stars   was  $10\kms$   in
spectrograph~1.

\begin{figure}
\ifthenelse{\UseFigs=1}{
\includegraphics[width=\hsize]{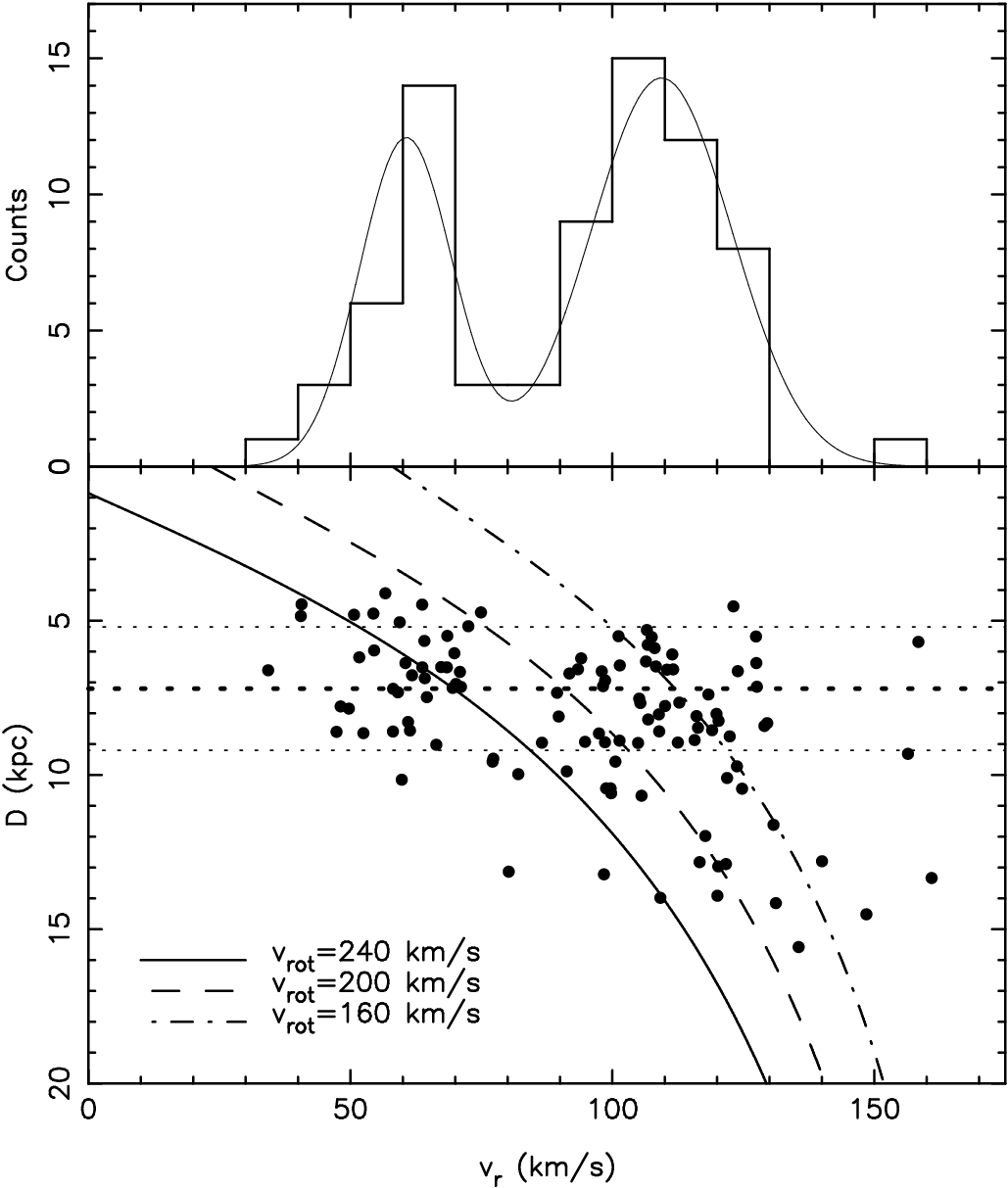} }{caption1}
\caption{The top panel shows  the distribution of radial velocities of
the M-giant targetted in our two  2dF fields (see text for details). A
clear bimodality  is present  that we fit  by a double  Gaussian model
(thin  line) using  a maximum-likelihood  algorithm.  We  identify the
second, more  numerous population  centred on $v_{r}=109\kms$,  as the
CMa overdensity.   The bottom panel presents the  expected phase space
behaviour of a  population of stars orbiting around  the Galaxy with a
mean   rotational   velocity   of   $v_{rot}=240\kms$   (full   line),
$v_{rot}=200\kms$  (dashed line)  or  $v_{rot}=160\kms$ (dotted-dashed
line).  The positions of the  M-giants have been overplotted as filled
circles and the mean distance of the CMa overdensity is represented by
the heavy dotted line at $D=7.2\kpc$. The light dotted lines represent
our distance selection criteria for the histogram of the top panel.}
\end{figure}

\begin{figure}
\ifthenelse{\UseFigs=1}{
\includegraphics[angle=270, width=\hsize]{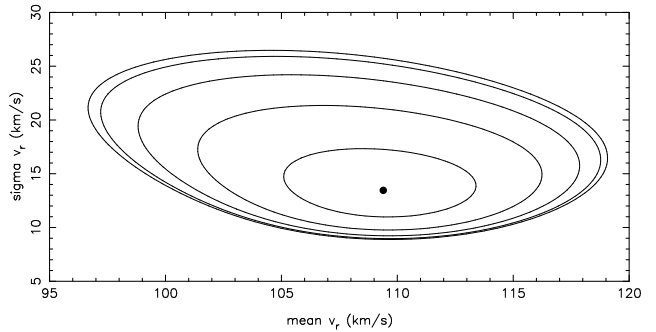} }{caption1}
\caption{Likelihood contours of the  mean and dispersion of the double Gaussian  model   fit  to  the  radial  velocity 
 data  displayed  on Figure~3. The  contours are spaced at $1\sigma$  intervals. The radial velocity distribution of the
 CMa population is well constrained, with a   mean   value   of   $109\pm4\kms$  and   a   dispersion   of
$13\pm4\kms$.}
\end{figure}

The  top  panel of  Figure~3  shows  the  distribution of  the  radial
velocities of the  75 M-giant stars present in  two 2dF fields centred
on  the  bulk   of  the  CMa  overdensity  ---   within  a  dergee  of
$(l,b)=(240.0\deg,-8.8\deg)$ and  $(l,b)=(240.0\deg,-6.8\deg)$ --- and
within $2\kpc$ of its  mean distance --- $5.2\kpc<D<9.2\kpc$. The most
striking   feature  is   the  bimodality   of  the   distribution.   A
Kolmogorov-Smirnov  test  shows  that   there  is  less  than  $0.1\%$
probability   that  this   distribution   is  drawn   from  the   best
single-Gaussian fit of the  data.  Therefore, a maximum-likelihood fit
of the sum of two normal distributions was performed and shows a first
peak centered at  $61\pm4\kms$ with a dispersion of  $9\pm3\kms$ and a
more populated  peak centered on  $109\pm4\kms$, with a  dispersion of
$13\pm4\kms$ (see Figure~4).  Taking the radial velocity uncertainties
into  account in  the maximum-likelihood  fit, as  estimated  from the
cross-correlation  peak width,  the  velocity dispersions  of the  two
peaks are $0\pm9\kms$ and $10\pm4\kms$, respectively.

Since the second peak contains almost twice as many stars as the first
and  since its  stars are  centered around  the distance  of  CMa (see
bottom panel of Figure~3), we tentatively identify it with the large CMa
overdensity  that appears on  Figure~2. Moreover,  if these  stars are
orbiting   the  Galaxy,   they   have  a   mean  rotational   velocity
($v_{rot}\sim160\kms$) that  is too low for  a thin disc,  and is even
low for a slower thick disc population ($v_{rot}\sim170\kms$, where we
assume a circular velocity of  $220\kms$ and take the thick asymmetric
drift  of  $-51\pm5\kms$  found  by \citealt{soubiran}).   Though  the
precise kinematic properties of the outer rim of the Galactic disc are
poorly known (and extrapolations from local measurements are likely to
be  misleading),  the  low  observed  velocity  dispersion  is  surely
incompatible   with  a   thick  disc   population,  which   has  Solar
Neighbourhood   velocity    dispersions   of   $(\sigma_U,   \sigma_V,
\sigma_W)=(63\pm6, 39\pm4, 39\pm4)\kms$ \citep{soubiran}.

The  position  and dispersion  of  the  first  peak are  difficult  to
explain. Indeed, its stars appear too far away to coincide with a thin
disc population contaminating  our sample which we would  expect to be
located  at  a distance  of  $D\sim3.5\kpc$  at  this radial  velocity
(assuming a  circular velocity of  $220\kms$). It is possible  that we
are significantly overestimating the distances to these stars; the fit
to  the  RGB we  determined  in  section 2  is  adequate  for CMa  but
populations  of  different  metallicity  and/or  age  could  follow  a
different relation.  However,  the very low dispersion of  the peak is
unexpected for  a disc population.  Another explanation  would be that
the CMa overdensity is composed of two populations with different mean
velocity.  The  low dispersion  of the two  peaks, similar to  what is
observed  in tidal streams  (see e.g.   \citealt{ibata97} for  the Sgr
dwarf or \citealt{ibata04} for the M31 stream), argues in favour of an
accretion scenario.

To summarize, the observed kinematics  of M-giants stars in the centre
of the CMa structure do not show a distribution that would be expected
for a single disc population or a Warp orbiting the Milky Way.

\section{Discussion and conclusions}

The clear asymmetry in addition to  the Warp in the direction of Canis
Major and the peculiar radial  velocity of these stars show that there
is an  extra population in  this direction, rotating around  the Milky
Way in a  prograde direction at a lower  mean rotational velocity than
what would be expected for disc  stars.  This structure at the edge of
the  disc  may  indicate  that  the  Galaxy shows  the  same  kind  of
substructure that were recently discovered  at the edge of the disc of
M31 \citep{ferguson}.

As we have already mentioned  in Paper~I, this CMa population could be
the  remnant of  the accretion  of a  dwarf galaxy  onto  the Galactic
plane.   With  a  mean  value  of  $v_{r}=109\pm  4\kms$,  the  radial
velocities of CMa  falls in the range of the  radial velocities of the
grouping  of  globular  clusters  identified  in  Paper~I  (see  their
Figure~12),\footnote{However, the prograde  simulation of an accretion
presented there is not compatible  with the radial velocity of the CMa
field since  it predicts a  velocity $\sim200\kms$ for the  CMa dwarf.
This is not surprising as  this explorative simulation was mainly made
to see if the accretion scenario could reproduce the observations, and
the kinematic information available at that time was meagre.} while it
also  correlates  well with  the  \citet{crane}  radial velocities  of
M-giants  belonging  to  the  ring-like  Galactic  Anticentre  Stellar
Structure. Moreover, the low  dispersion of the distribution of radial
velocities of  CMa M-giants is compatible with  an accretion scenario.
While this could be a sign that the `Ring' and the CMa overdensity are
related,  it should  also be  noted that  the distances  to  these two
features are different.  Therefore, it may not prove  possible to link
the  two in  a direct  way and  would require  an  alternate accretion
scenario, in  which the  CMa dwarf  has been in  the process  of being
accreted for a longer time  than previously thought.  In this case, it
would not have created the `Ring'  and the CMa overdensity in the same
passage around the Milky Way  but in successive orbits, with its tidal
arms wrapped  a few  times around the  Galactic disc.   If simulations
reveal this is  a plausible scenario, it could  also be interesting to
see    whether    the   low    latitude    structure   presented    by
\citet{rocha-pinto}  in  the  Triangulum-Andromeda  direction  can  be
explained in this way or whether it is too far away from the Galaxy to
be created by the same accretion process.

Finally and with the recent discovery of a putative distant spiral arm
in (mainly) the  fourth quadrant of the Milky  Way \citep{mcclure}, it
is worth  considering the possibility  that the CMa overdensity  is in
fact the prolongation of this  spiral arm to lower longitudes. Even if
their observations  do not  overlap with CMa,  the radial  velocity we
present  here could  be compatible  with their  velocities  that reach
$v_{r}\sim110\kms$  at  $l\sim255\deg$.  The  age  of  the  CMa  stars
(Paper~II)  and  the  structure  of the  overdensity  (Paper~I)  would
however be hard to explain in this scenario.

The confusion  between the  CMa overdensity and  the Warp  that caught \citet{momany} brings  up questions on the real 
nature and dimensions of the  Warp. How much  of the Canis  Major overdensity has  until now been taken as  the Warp?
Disentangling the role and extent of each may prove to be a difficult task.

\section*{Acknowledgments} NFM would like to thank Annette Ferguson for useful discussions on the Galactic warp. The
referee is thanked for useful comments that helped improve the letter.

\newcommand{\mnras}{MNRAS}
\newcommand{\nat}{Nature}
\newcommand{\araa}{ARAA}
\newcommand{\aj}{AJ}
\newcommand{\apj}{ApJ}
\newcommand{\apjl}{ApJ}
\newcommand{\apjs}{ApJSupp}
\newcommand{\aap}{A\&A}
\newcommand{\aaps}{A\&ASupp}
\newcommand{\pasp}{PASP}

\end{document}